\documentclass{appolb}
\usepackage{graphicx}

\usepackage{type1cm}        
\usepackage{makeidx}         
\usepackage{graphicx}        
\usepackage{multicol}        
\usepackage[bottom]{footmisc}

\usepackage{newtxtext}       %
\usepackage{newtxmath}       
\usepackage{slashed, mathtools} 

\newcommand{\G}{\mathcal{G}}

\newcommand{\be}{\begin{equation}}
\newcommand{\ee}{\end{equation}}
\newcommand{\ba}{\begin{eqnarray}}
\newcommand{\ea}{\end{eqnarray}}
\newcommand{\bi}{\begin{itemize}}
\newcommand{\ei}{\end{itemize}}
\newcommand{\la}{\label}
\newcommand{\non}{\nonumber}

 \newcommand{\eela}[1]{\quad\hbox{\scriptsize{#1}}\label{#1}\end{eqnarray}}
 \newcommand{\eelb}[1]{\label{#1}\end{eqnarray}}

        \def\be{\begin{eqnarray}}    \def\ee{\end{eqnarray}}
 \def\bi#1{\begin{itemize}\item[#1]}     \def\ei{\end{itemize}}  
   \def\^#1{\hat{#1}}

 \def\a{\alpha}         \def\g{\gamma}      \def\G{\Gamma}
 \def\d{\delta}         
       \def\l{\lambda} \def\L{\Lambda}     \def\m{\mu}
                 \def\n{\nu}
             \def\r{\varrho}     \def\s{\sigma}  
 \def\t{\tau}          
               
 \def\w{\omega}

 \def\ffract#1#2{\raise .2 em\hbox{$\scriptstyle#1$}\kern-.3em/
                 \kern-.2em\lower .15 em \hbox{$\scriptstyle#2$}}

\def\bmatrix{\begin{matrix}} \def\ematrix{\end{matrix}} \def\bpmatrix{\begin{pmatrix}}\def\epmatrix{\end{pmatrix}}
\def\bcenter{\begin{center}} \def\ecenter{\end{center}}

\begin{document}
\title{Minimalistic musings about the Standard Model}

\author{Chris P. Korthals Altes 
\address{NIKHEF Theory Group, NIKHEF Amsterdam. Science Park 105, 1098 XG Amsterdam and
Centre de Physique Th\'eorique, Campus de Luminy, 163 Avenue de Luminy, 13009 Marseille }

}

\maketitle
\begin{abstract}

\hspace{2,0cm}{{\bf A tribute  to Martinus Veltman 1931-2021}}\\

 Martinus ("Tini") Veltman 's early contributions to the Standard Model  were essential for its succes. After some nostalgic reminiscences, I turn  to the   Standard Model with a minimalistic attitude, the point of view that beyond the SM there is only the Planck scale. Known since long, the gravitational force can be obtained as the  gauge  theory of  local Poincar\'e symmetry, called gauge gravity. 
  This gauge theory of gravity embodies {\it per se} a Palatini formulation. 
 This causes the potential of non-minimally coupled Higgs inflation  to have  an intriguing  improved large field behaviour. Some of its effects are experimentally accessible or refutable. The  question of quantum corrections is discussed.  

\end{abstract}
  
\section{Introduction}

 In writing a tribute to Prof Tini Veltman  it occurred to  me that 
 he was like many a physicist prey to periods of elation ranging to bouts  of despair about perceived lack of understanding of his work. 
 He had certainly some reasons to have  despaired, when in the late sixties he was analysing gauge theories for weak interactions, which were, so he was told by colleagues, "one of those funny corners of physics".

 But in Tini's case  elation had always the last word, amongst more as recipient at the 1999 Nobel banquet in Stockholm's town hall. 
  His elation did flow over to the audience  in his after dinner speech; he likened the anguish of potential  Nobel candidates all over the world,  waiting for the call from Stockholm, to the terror spread by the Viking ancestors of the Nobel committee a millennium ago  amongst innocents on European shores.

Indeed it is with due circumspection  that I write these lines. Tini would have despaired about  the title and abstract.

Actually he had, as far as I know, four famous encounters with gravity that went into print. 
The first one was in a paper with Henk van Dam~\cite{tinihenkvd} on the lack of a continuous limit from massive vector gauge theory and Einstein gravity theories to their massless and concluded that the latter should be strictly massless.
The second one on the one loop gravity
action~\cite{tinioneloop}, with Gerard 't Hooft. The third one was a Letter on the effect of the Higgs condensate 
as source of the gravitational field  brought up by Linde~\cite{tinicosmohiggs} a few months before.  The fourth one in his lecture course on gravity at the Les Houches school\cite{tinileshouches}.

The first  and second papers are quite influential till the present day. 

His lectures are typical for the particle physicist he was, in heart and soul. He did forego
the geometrical approach, and presented gravity as the theory of gravitons. The lectures 
are reflecting the point of view taken  taken in the CERN report "Diagrammar" ~\cite{diagrammar}.
That is, for him the quantum theory of gravity was the collection of diagrams. This mantra had served him well in the elucidation of radiative corrections to weak and electromagnetic interactions. 

Curiously, he did not appreciate the usefulness of solitons or instantons in theories like QCD.  Not only in QCD.  I was present when he  explained to a well known cosmologist
in the CERN coffee room why black holes cannot possibly form- although he may well have played the devil's advocate, something he loved to do in order to challenge people.  Not only solitons but also certain disciplines in physics did not 
carry his approval.  At lunch with a group of condensed matter physicists in  a  restaurant in the Quartier Latin, during a meeting at the Ecole Normale, he aired the opinion
that Statistical Mechanics was the art of finding an interesting physics property  and then averaging it out.

I had a brush with him in the corridor of CERN just after the second gravity paper appeared, because I asked him why gravity had anything to do with particle physics. Why was gravity quantized? After all gravitons had not been seen. 
After starting of course a muscular response, to my astonishment he acquiesced.  Nowadays astrophysical black holes have been detected, but the discussion how to detect effects from the quantum nature of gravity~\cite{azumdar} and detect gravitons~\cite{dyson} is very interesting, but  still at an academic  stage. And Tini knew very well that quantum gravity,  with his approach or any other, was  a formidable opponent  and he did not engage it anymore after the mid-seventies, at least not in published form. 
 
Tini entered my horizon when he gave a seminar in the mid-sixties in Amsterdam where I studied theory.  He wore sandals, a pullover and  displayed a very caustic sense of humour.  This in stark contrast to the professors  in Amsterdam  who wore- at that time at least- a  conventional outfit.  

A little later, involved myself in current algebra, I profited from  his PRL letter on the divergence equations~\cite{divergenceeqn}. It explained wonderfully well how sum rules  of current algebra could be obtained from low energy theorems and dispersion relations, avoiding any discussion of Schwinger terms. This paper signalled the start of his conviction  that non-abelian vector fields were the key to understanding weak interactions.

Tini was a regular guest at the CPT in Marseille, from 1968 on, partly because the man in charge there, Toni Visconti, and his students were  early birds  in computer generation of Feynman graphs and wanted to profit from Tini's "Schoonschip" expertise.  I had arrived there as a postdoc after  the 1971 revolution at the Amsterdam HEP conference, where  Tini had Gerard 't Hooft present part of his thesis work~\cite{thooft71}. 

At the 1971 Orsay Summer Institute Tini told me that he worried about the chiral anomalies in what was then still the purely leptonic version of the SM.  Michel Perrottet and I wrote a paper belabouring this worry~\cite{perrottet}.  In 1972 both Tini and Gerard played a crucial role at the Marseille conference on Yang Mills fields (at that time the Standard Model including quarks was two months old~\cite{ilio}) and on a similar occasion in 1974. 

Of course the famous Summer Institutes at the Orsay, Ecole Normale and Triangular Meetings were hotbeds for developing Particle Theory and Tini and Gerard were faithful participants, with crucial contributions. 

 In 1974 Tini had invited me to spend some months in Utrecht, which happened to coincide with the advent of the 1974 October revolution. He discussed his paper
on the Linde effect extensively before he sent it off.  But the ongoing revolution posed more pressing subjects....

From autumn 1978 till spring 1981 I spent more than two years in Utrecht. I had decided to work on numerical lattice gauge theory although Tini certainly did not encourage me to do so. He was busy with his work on  radiative corrections in the electro-weak sector and gave his well-known talk ("Quarks and Leptons: what next?") at the 1979 Lepton Photon Conference at Fermi Lab.. It makes interesting reading still  now. Bottom was then just discovered. Tini discussed  the expected top-bottom mass difference in connection with custodial symmetry and radiative corrections to the $\rho$ parameter.  He was playing with a Higgs heavier than the unitarity limit of vector boson scattering, creating a strongly interacting sector. But his talk ends with the admonition "We should keep checking electron muon universality".  

Well, his advice  was heeded, according to the latest  news from LHCb~\cite{lhcb}.

Some years later he spent   long periods  at the Theory Department of the  Universita Autonoma in Madrid, working with Francisco ("Paco") Indurain.
I happened to have an ongoing collaboration at that very same place and time with Antonio Gonzalez Arroyo. Apart from enjoying a marvellous hospitality
I had ample time to observe the  flamboyant encounters that Tini had with the quite picturesque population of the
Residencia de Estudiantes, a revered Madrile\~no landmark where we had the honour to stay.  It had traditionally harboured many  intellectuals and scientists from in and outside Spain.

After these recollections I would like to come back to the minimalistic view of the Standard Model, the fact that gravity
is just like the electroweak and strong force a gauge force, and how it may embody inflation.   

Of course  this idea of "gauge gravity" is an old idea, starting with a series of papers precisely sixty years ago~\cite{kibble}.  

 Kibble's  conclusion\footnote{For an identical view see also Weyl~\cite{weyl1950}} was that 
the theory was for all practical purposes indistinguishable from Einstein's theory.  More precisely, in absence of matter they are identical, but in the presence of fermionic matter they are differing  by a cosmological term 
\be
\sim\sqrt{-g} M_p^{-4} \left(\sum_f\bar\psi(x)\g^a\g^b\g^c\psi(x)\right)^2.
\ee

Here $M_p=\sqrt{(8\pi G_N)^{-1}}$ is the reduced Planck mass, and the indices $a,b,c$ on the Dirac matrices are fully anti-symmetric. The sum is over all fermions in the SM, provided the right handed
neutrinos are Dirac type. The square of Newtons constant renders all effects of this term unobservably small.

However if one introduces a non minimal coupling of the Standard Model  Higgs added to  the curvature scalar ${1\over 2}M_p^2 R$,
\be
\xi H^\dagger H R, 
\ee
\noindent then the behaviour of the Higgs potential for large field values is crucially changed, with measurable consequences in inflation. This idea, very much in the spirit of minimalism, is due to Bezrukov and Shaposhnikov~\cite{misha2008} in the context of Einstein gravity.

Subsequently Bauer and Demir~\cite{bauer} realized that  the Palatini formulation~\cite{palatini} of Einstein gravity improved parametrically (in terms of $\xi$) this inflation potential for high values of the Higgs field. Recall that the Palatini method starts from positing the metric and the   connection  to be independent  variables in the Einstein Hilbert action. Varying the action with the connection gives an equation of motion which is nothing but the metricity condition  on the metric, i.e. that the metric is covariantly constant. 

And gauge gravity has a Palatini method  built in, essentially because  the local translations and the local Lorentz group  generate independent connections, as will be amply clear from Section (\ref{sec:kibble}). 

Controversy sets in when quantum effects are taken into account.   The traditional approach is to look for symmetries  that are respected on the quantum level. That tells you the general structure of the effective action for the Higgs .  The crucial symmetry is the approximate translation invariance of the asymptotically flat Higgs potential.

  I was ruminating on this  idea  in late summer 
 last year, thinking that gauge gravity was a fairly remote and quiet corner. Till I discussed with Jan Smit who soon brought to my attention a couple of quite recent papers by Misha Shaposhnikov and coworkers~\cite{misha2020}.  So  what follows is mostly history and I do sincerely apologise for incomplete referencing.  
 
 How gauge gravity emerges from local Poincar\'e  invariance is narrated in Section (\ref{sec:kibble}). I couldn't withstand the temptation to tell this old story
 in the form of  a pastiche using two well known personalities. Both came into existence practically  on the CERN site,  in Ferney-Voltaire two and a half centuries ago
\footnote{Due to  Voltaire's fulminations~\cite{voltaire}  against  Leibniz' idea of "All is for the best in the best of all possible worlds"}.
 They are the legendary  Candide and his companion, Dr Pangloss. They are supposed to have learnt how the forces in  the standard model are obtained by the gauge principle but never learnt about  any theoretical description of gravity, in particular they are blissfully unaware of use of geometrical ideas. 
However they  are  supposed to be acquainted with the experiments that infer that all bodies fall equally fast {\it in vacuo}.  This permits them to almost deduce that the force associated with the local Poincar\'e group is that of gravity.  The pastiche serves a dual purpose, to realize how simply  the SM begets  gauge gravity and how much we are preconditioned by the beautiful geometric point of view.

In Section (\ref{sec:higgsinflation}) I discuss Higgs inflation in the light of gauge gravity.

\section{In which Candide contemplates  gauge gravity potentials}\la{sec:kibble}

 As already mentioned in the introduction the strict believer in the gauge principle of the SM is Dr Pangloss. He is very intent on finding out what physical force is associated to the gauging of Poincar\'e symmetry. Unfortunately he is prone to pedantry and condescendence. Candide  has his reservations and is waiting to punch on weaknesses in the expos\'e of his colleague.

The two  decide to write a  Poincar\'e transformation as 
\be
{x'}^\m=\L^\m_{~\n} x^\n+a^\m,~~~ \L^\m_{~\n}\eta^{\n\r}\L^{~\s}_\r=\eta^{\m\s}, \eta=(1,-1,-1,-1).
\la{poin}
\ee
Dr Pangloss'  aim is to introduce a gauge covariant derivative  for the Poincar\'e symmetry made local. This asks for gauge potentials.  Candide concurs to take the potentials $t^\m_a$ for the translations.
with $\partial_\m$ as the translation charge. The index $a=0,1,2,3$. 
The local Lorentz symmetry needs a potential $A_{ab\m}$ with the normalized Lorentz group generators $M^{ab}$ as charges. The latter will depend on whether one has a quark ($M^{ab}=\g^{[a}\g^{b]}/2i$) or a vector boson ($(M^{ab})_{cd}=\d^{[a}_c{\d^{b]}}_d)$). They should appear in covariant derivatives.

They decide to  find the covariant derivative for  the Higgs field  $H$ first, as that should be the simplest case.
If the transformation is global    $H^{\prime}(x^{\prime})=H(x)$ and
\be
\partial^{\prime}_\m H^{\prime}(x^{\prime})=\L_\m^\n\partial_\n H(x).\non
\ee 
 want a derivative $D_a$, such that  in case ${x^\prime}^\m=\L^\m_\n(x)x^\n+a(x)^\m$
\be
D^\prime_a H^\prime(x^\prime)=\L(x)_a^b D_b H(x).
\la{covpoin}
\ee

First they note that the local Poincar\'e transformation is a general coordinate transformation, and $H'(x')=H(x)$
 as befits a scalar. Hence 
\be
\partial^\prime_\m H^\prime(x^\prime)={\partial {x}^\m\over{\partial {x^\prime}^\n}}\partial_\n H(x).\non
\ee
However this is not yet what they want, see Eq. (\ref{covpoin}).  But if they put $D_a=t(x)_a^\m\partial_\m$ then assuming
for the translation potential the rule
\be
{t^\prime}_a^\m(x^\prime)={\partial {x^\prime}^\n\over{\partial {x}^\m}}\L_a^b t(x)^\n_b,
\la{vierbeintransf}
\ee 

\noindent their Eq. (\ref{covpoin}) results.   
 
 The Higgs kinetic term  becomes $\int d^4 x~ t~ (D_a H)^\dagger D_b H\eta^{ab}$, $t$ being the determinant of the inverse  $t_\m^b$ of $t(x)^\n_b$.\footnote{Plugging in the definition of $D_a$  gives  the minimal coupling of a tensor $g^{\m\n}\equiv \eta^{ab}t(x)^\m_a t(x)^\n_b$ to the usual derivatives. From this identification follows that the translation potential is nothing but the Vierbein. But remember that our friends are unaware of metric, connections etc..  .} 
 
 Candide makes the point that  the transformation law (\ref{vierbeintransf}) does not have an inhomogeneous term, whereas the transformation law for the local Lorentz potential has one,  by construction.  He suggests another name to  avoid confusion
 and he proposes the name Vierbein for $t_\m^a$.

 Dr Pangloss agrees and proposes  to carry on with the less simple  fermion case with $\psi^{\prime}(x^\prime)=S(\L)\psi(x)$. 
 
 Were it not for the transformed coordinate
 on the left hand side they  would be concerned with an  internal symmetry operation as in the standard model with an $SO(3,1)$ potential
 $A(x)_{ab\m}$, and a familiar covariant derivative
 
 \be
 D_\m(A)=\partial_\m+iA_{\m},~~ A_\m={1\over 2}A_{ab\m}M^{ab}.
 \la{covderA}
 \ee
 
 This derivative transforms under a local Lorentz transformation as they are used to in the standard model to obtain a covariant derivative. 
And after a local Poincar\'e transformation $x^\prime(x)$ in $\psi(x^\prime)$ it behaves as a contravariant vector 
\be
D^\prime_\m(A^\prime)\equiv\partial^\prime_\m+iA^\prime(x^\prime)^\prime_{\m}={\partial x^\n\over{\partial {x^\prime}^\m}} D_\n(A).
\la{Amucovariant}
\ee

So they obtain for the combination 
\be
D^\prime_\m(A^\prime)\psi^\prime(x^\prime)={\partial x^\n\over{\partial {x^\prime}^\m}}S(\L)D_\n(A)\psi(x).\non
\ee
Like for the scalar case this is not yet their desired result. They multiply the gauge covariant derivative (\ref{covderA}) with the vierbein $t^\m_a$ to get for the local Poincar\'e covariant derivative $D_a(A)=t^\m_a\partial_\m+i t^\m_a A_\m$
\be
D^\prime_a(A^\prime)\psi^\prime(x^\prime)=\L(x)_a^b S(\L)D_b(A)\psi(x).\non
\ee
Dr Pangloss  notes that $D_a(A)$ is the sum of a translation term $t^\m_a\partial_\m$ and a local Lorentz transform
 term $t^\m_a A_\m$. This is as expected from a local Poincar\'e invariant derivative. However as brought up by Candide already in the scalar case  the translational  "potential" has no inhomogeneous term.

The kinetic term  for the fermion field becomes in terms of the new covariant derivative
\be
L_f(\psi,D_a\psi)=\bar\psi i\g^a D(A)_a\psi. \non
\la{dirackin}
\ee

However Candide demurs. He repeats his observation made in the scalar case. Only the local Lorentz potential has an inhomogeneous part. He then goes on reproaching Dr Pangloss that he has produced a a nice mathematical framework to obtain a local Poincar\'e invariant action for scalar and fermion fields  but no physical interpretation.  Where is the potential with an inhomogeneous term under a general coordinate transformation? 

The two gentlemen agree to to disagree and to reconvene quickly.


\subsection{In which Candide discovers what describes the gravitational force}

The discussion is taken up again with Dr Pangloss announcing he has found an answer to the question of Candide.
He claims to have found a combination of the translation (i.e. the Vierbeins) and local Lorentz potentials that has the property 
Candide desires.

In what follows Dr Pangloss is explaining himself within the quotation marks.

"Like in the standard model the commutators of our derivatives give the field strengths. In our case I have two.
One is the field strength formed from
\be
[D_\m(A), D_\n(A)]\equiv{1\over 2} \hat R(A)_{\m\n}.
\la{so31fieldstrength}
\ee

This is the $SO(3,1)$ gauge  field strength and can be written in component form $\hat R_{\m\n}=\hat R^{bc}_{~~\m\n}M_{bc}$. I can multiply  this  tensor with the Vierbeins $t^\kappa_b t^\l_c$ and the result is the tensor $\hat R^{\kappa\l}_{~~\m\n}$.

On the other hand I can form the commutator from the derivatives $D(A)_a=t^\m_a D(A)_\m$.  The result is {\it different} from $\hat R_{\m\n}t^\m_a t^\n_b\equiv \hat R_{ab}$
\ba
[D_a, D_b]\psi &=&{1\over 2}\hat R_{ab}\psi- T^c_{ab}D_c\psi\non\\
 T^c_{~ab}&=&-\left(t_a^\m t_b^\n- t_b^\m t_a^\n\right)D_\m(A)t_\n^c.
 \la{torsion}
\ea
The torsion $T^c_{~ab}$ is the new element, obviously antisymmetric in the last two indices\footnote{The action of $D(A)_\m$ on the Vierbein  is defined as
\be
D_\m(A)t_\n^c=\partial_\m t_\n^c+A^c_{~d\m}t^d_\n.\non
\la{DAvierbein}
\ee}
. Let me project it on Einstein indices $T^\r_{~\s\t}$ with the Vierbeins . It is then the  antisymmetric
part of a new quantity, a composite of Vierbeins and $SO(3,1)$ gauge potential,
\be
 \G^\r_{~\s\t}(x)\equiv -  t^b_{~\s}D_\t( A)t_{~b}^\r . "
\la{affineconn}
\ee

At this point Candide gets impatient and asks what all this is good for. Dr Pangloss implores him to let him finish. 
   
He shows how  $\G$  transforms under general coordinate  transformations by using in (\ref{affineconn}) the transformation properties (\ref{Amucovariant}) and (\ref{vierbeintransf}).  He comes up with the following formula for the transformation law 
\be
{\G^\prime}^\r_{~\s\t}(x')= {\partial {x^\prime}^\r\over{\partial x^\l}}{\partial x^\m\over{\partial {x^\prime}^\s}}{\partial x^\n\over{\partial {x^\prime}^\t}}\G^\l_{~\m\n}(x)+{\partial {x^\prime}^\r\over{\partial x^\m}}{\partial^2 x^\m\over{\partial {x^\prime}^\s\partial{x^\prime}^\t}}.\non
\ee

  Candide is delighted  by this expression: "Dr Pangloss, this is a very interesting finding. Pray, let us try to interpret it."
  
  He starts by  noting that this law permits the symmetric part of the new quantity $\G$ to be transformed away, at least locally, where the first derivatives can be set to unity. Dr Pangloss  concurs in this view. The latter then remembers the experiments of  his colleagues  Philoponus~\cite{philoponus} and Galileo~\cite{galilei} that indicate all bodies fall equally fast
 in vacuo. Candide  argues that this is consistent with  the property  just established. 
 
 "Alas, my friend," answers the inexorably pedantic Dr Pangloss, "you may be too quick.  The experiment shows all gravitating bodies feel the same acceleration. But    
 a uniform acceleration of the system may undo the gravitational one only if the inertial mass and the gravitational mass
 are the same." Candide has to give in, and says they will have to wait for the outcome of the relevant experiment. 
 
 Nevertheless the two gentlemen are quite satisfied with their result. Pending a positive experimental outcome  the gravitational force can be described by this  composite gauge  potential $\G^\kappa_{~\l\m}$.  
 
 They embark now on a natural question. What is the field strength related to this gauge potential?
   Dr Pangloss quickly comes up with the answer 
 \be
R(\G)^\kappa_{~\l\m\n}=\partial_\m\G^\kappa_{~\l\n}+\G^\kappa_{~\a\m}\G^\a_{~\l\n}-\m\leftrightarrow \n.\non
\ee

They do check that after a general coordinate transformation indeed the inhomogeneous part  cancels   as behooves a field strength.

As usual  Candide gets uneasy.  "Dr Pangloss, should this gravitational field strength not be related to our local Lorentz field strength (\ref{so31fieldstrength}) ?" 

The latter does agree that this may be true. The answer he gives is obtained by plugging his original definition
 of $\G$ in terms of  translation and local Lorentz potential, Eq. (\ref{affineconn}), into the new field strength.
 
 And indeed, Candide's expectation was justified. The field strength of gravity turns out to be a contraction of the SO(3,1) field tensor with two Vierbeins

\be
\hat R( A)_{ab\m\n}t^{a\kappa}t^b_{~\l}=R(\G)^\kappa_{~\l\m\n}.
\la{Riemannso31}
\ee

The two are quite happy with their findings. The field strength of gravity $R(\G)$ is identical to the SO(3,1) field strength contracted with two Vierbeins.  They conclude  that the gravitational force is described by a gauge potential, a composite of Vierbein fields and the SO(3,1) gauge potential. They decide for a good dinner and exit the narrative.

\subsubsection{Epilogue to the dialogue} 

Clearly, had the hands of the protagonists not been tied up on their
back by their lack of knowledge of geometric concepts the whole conversation could have ended at Eq. (\ref{affineconn}). The latter becomes after some trivial rearrangement the condition of metricity on  the Vierbein field.

The geometry for an affine connexion with torsion was developed by Cartan~\cite{cartan}. In the seventies Cartan's geometric view point was taken up again by Trautman and Hehl~\cite{cartan}. 

Still we have to choose the pure gravity term in the SM action. Of course there  is the curvature $R(\G)$. But with torsion the Riemann tensor is no longer 
symmetric in the exchange of the first pair of indices with the last pair. Hence the  pseudo scalar  curvature $\epsilon.R$ does not vanish. There is one more independent  scalar that can be taken as the Nieh-Yan invariant~\cite{misha2020}. In this note we will
stick to the curvature.

What is certainly not academic is  that this gauge gravity has a Palatini type formalism built in through
the local Lorentz potentials. They are independent of the Vierbein fields. Just like in Einstein gravity the Palatini point of view is the independence of the affine connection of the metric.

In conclusion the Standard Model produces gauge gravity, along with the strong and e.m. forces. By constraining the torsion in Eq.(\ref{torsion}) to vanish one retrieves Einstein's gravity. 

 And luckily, the original hypercharge  gauge anomaly cancellations of  including the strong,  electroweak and hypercharge forces~\cite{ilio}  do include as well that of gravity~\cite{marshak}!  They fix the hypercharge assignments, once the electric charge, say of the electron, is fixed.

Evidently there is no clue why the electroweak and gravitational mass scale are so far apart. 
That is, scale invariance is badly broken. 

Which brings us to  Weyl transformations.  They do replace the scaling of coordinates and become an {\it internal} symmetry when the Vierbein fields are available.   They  do act multiplicatively on Vierbeins.  But they do leave the Lorentz potential invariant. This is possible because the latter is independent of the Vierbein- like in the Palatini approach to Einstein gravity the connection is independent of the metric tensor. 

 Hence the covariant derivative $\partial_\m +i A_\m$ is invariant. And so is their commutator $\hat R(A)_{\m\n}$ in (\ref{so31fieldstrength}).  Hence the Riemann curvature  transforms according to Eq. (\ref{Riemannso31})  like the product of two  Vierbein fields 
\footnote{On matter fields they act with a power fixed by the engineering dimension. Symmetry under Weyl transformations admits only dimensionless couplings.}.  
There is good use for this  in section (\ref{sec:higgsinflation}) when discussing the high field behaviour of the Higgs field in Higgs inflation.

\section{Does the Standard Model accommodate the Inflationary Universe?}\la{sec:higgsinflation}

Inflation refers to a period in the early universe where the energy of the universe  was all stored in the potential energy of the  inflaton field.   In order to have 
sufficient inflation  the field is supposed to roll slowly, i.e. the potential energy
of the field at these high values should be flat, i.e. approximately translation invariant.  

This flatness obviously   excludes the Higgs potential $V_{ew}={\l\over 4}\left(H^\dagger H-{v^2\over 2}\right)^2$ with $ v\sim 246 GeV$  as a candidate for inflation.  

Bezrukov~\cite{misha2008} et al. exploited the fact  that Newtons constant decreases or rather  the Planck mass  increases when they added the Higgs coupled to curvature

\be 
M_p^2\rightarrow M_p^2+2\xi |H|^2 .\non
\ee

To see the consequences  at high field values  the traditional approach is  to make a Weyl transformation on  the gravitational field such that  in the new, "Einstein" frame the constant in front of the curvature is again $M_p^2$ and the Higgs field in that frame has a canonical kinetic term~\footnote{For a justification see reference~\cite{unruh}}. The Higgs
potential in the Einstein frame then becomes due to the decrease of Newton's constant
\be
\hat V(H^\dagger H)= {V_{ew}(H^\dagger H)\over{(1+2\xi {H^\dagger H\over{M_p^2}})^2}}.
\la{renormbyweakgravity}
\ee
 
 Now the potential is flat  at high field values $|H|>>{M_p\over \sqrt{\xi}}$ whereas the electroweak breaking at $|H|\sim v$ is left unchanged unless $\xi\sim 10^{17}$.   
 
 If we want the inflationary plateau large enough to create enough e-foldings $\xi$ must be large, on the order of $10^4$ ($10^{7-8}$ in the Palatini case).  The original $raison~ d'\^etre$ of the $|H|^2R$ term was that it enhances an internal symmetry, Weyl symmetry, for a fixed numerical value of the coupling~\cite{callan}, $\xi=-1/12$.  A huge value of the coupling seems  unnatural. However we want to describe inflation, an event where scale invariance was broken in an unprecedented fashion.  It is then somewhat of a miracle that a numerical window of couplings describing successful inflation exists at all.

  To see in more detail how the  flattening of the potential comes about we  write down the relevant part
 of the action in terms of the radial Higgs field $h$ with $h^2=2 H^\dagger H$ and drop its couplings to the gauge fields
 \be
 S_J=\int d^4 x ~\sqrt{-g}\left( -{1\over 2}(M_p^2+\xi h^2 )R+{1\over 2}g^{\m\n}\partial_\m h\partial_\n h -
 V_{ew}(h)\right).\non
 \la{jordan}
 \ee
 This is the action in the Jordan frame. It has a Higgs field dependent gravitational coupling and the canonical kinetic term for the Higgs and the canonical potential causing electroweak breaking.  The question is what the effect of the Higgs dependence of the gravity coupling will be on the potential.  This can be computed directly~\cite{unruh} in the Jordan frame. 
 
 Alternatively a Weyl transformation $\exp(-\w)$ on the gravitational field can restore the coupling in front of the curvature by the physical coupling  $M_p^2$~\footnote{The discussion here is in the context of gauge gravity, but the mathematics is identical to that in ref.~\cite{bauer}}.  This Weyl transformation acts only on the 
 Vierbein fields, and using  the relation between the curvature tensor and the local Lorentz
 field strength we see that the curvature $R$ and the metric transform both like $\exp{(-2\w)}$ and the determinant  like $\exp{(4\w)}$
 \be
 S_E=\int d^4 x ~\sqrt{-g}\left( -{1\over 2}(M_p^2+\xi h^2 )\exp{(2\w)}R+{1\over 2}g^{\m\n}\exp{(2\w)}\partial_\m h\partial_\n h-
 \exp{(4\w)}V(h)\right).\non
 \la{einstein}
 \ee
 This is the action in the Einstein frame.
 Requiring that the curvature coupling is the Planck mass fixes $\w$ in terms of $h$
 \be
 \exp{(-2\w)}=1+\xi {h^2\over{M_p^2}}.\non
 \ee
 The prefactor $\exp{(4\w)}$ of the Higgs potential is thereby determined 
 as in  Eq. (\ref{renormbyweakgravity}).

 The Higgs kinetic term is normalized in the Einstein frame by introducing the renormalized Higgs $\phi$.   It becomes quite simply related to $h$ by
 \be
  h={M_p\over{\sqrt{\xi}}}\sinh\left(\sqrt{\xi}{\phi\over{M_p}}\right).
 \la{palatinihphi}
 \ee 
 So in the Einstein frame  the field with a canonical kinetic term is very non linear function of the original Higgs field. 
 And the potential $U(\phi)=\hat V(h(\phi))$ becomes

 \be
 U_{Pal}(\phi)=U_\infty \tanh^4(\xi \phi/M_p^2).
 \la{palatiniphi}
 \ee
 
 The potential has an asymptotic translation symmetry in Einstein frame. This what we need for inflation. In the Jordan frame this
 becomes an asymptotic scale symmetry. 
 
 So far for  gauge gravity.
 
 For Einstein gravity the torsion in Eq. (\ref{torsion}) is vanishing.  So the Lorentz potential depends now on the Vierbein and 
 so does the affine connection in (\ref{affineconn}). Hence we have to do with metric Einstein gravity, not Palatini-Einstein gravity.
 
 As a consequence the simple multiplicative renormalization of the Higgs kinetic term  gets  replaced by 
 \be
 \exp(2\w)\rightarrow \exp(2\w)+\exp(4\w)\xi^2h^2/6M^2_p.
 \la{dim6}
 \ee 
 Note that for small $h/M_p$ the new term is a dimension 6 operator suppressed by a factor $\L_m=\sqrt{6}M_p/\xi$.  
 
 The relation of $h$ to the renormalized Higgs field $\phi$ is for large fields and large value of $\xi$ 
 \be
 h={M_p\over\sqrt{\xi}}\exp({\phi\over{\sqrt{6M_p}}}),  \mbox{metric Einstein gravity}.\non
  \ee
 and the potential in the Einstein frame becomes under the same conditions 
 \be
 U_m(\phi)=U_{\infty}\left(1-2 \exp{\left(-2\phi/(\sqrt{6}M_p)\right)}\right).
 \la{metrichphi}
 \ee 
 
 For  slow roll the inflationary  Hubble parameter equals
 \be
 H_I^2={U_\infty\over 3 M_p^2}={\l\over 24}\left({M_p\over{\xi}}\right)^2,
\la{H_I}
 \ee
 \noindent  and this will be the bench mark scale for the inflationary regime. 
 
 With the input of the potentials (\ref{palatiniphi}) and (\ref{metrichphi})  can extract the  values of  $\xi$  and other inflationary parameters
 and compare them in both cases.

 Inflation in the Einstein frame is taking place between $\phi_i$ at time $t_i$ and $\phi_f$ at $t_i$ and
 the number $N$ of e-foldings in that frame the integral is given by the integral
 \be
 N={1\over {M_p^2}}\int_{\phi_f}^{\phi_i} U(\phi)/\left(\partial U(\phi)/\partial\phi\right) d\phi.\non
 \ee
  
We skip further details on the traditional comparison to data  and give the results.
The upshot  is that the number of e-folds and the scalar tilt are numerically pretty much the same in the gauge gravity case and in the Einstein gravity case~\cite{misha2008}. However the $\xi$ value for gauge gravity is $10^{7}$ while $10^{4}$ for Einstein gravity. It follows that  for gauge gravity the ratio tensor to scalar tilt is very small, $\sim 10^{-11}$ for gauge gravity, seven orders of magnitude smaller than its Einstein gravity value.  

Unfortunately the measurement of 
this ratio will be very hard below $r\sim 10^{-5}$. Of course, it suffices to find an experimental value for $r$ above 
this barrier and the Palatini case is falsified.

The current upper bound ~\cite{planck} is $r<0.064$ and together
with the Hubble rate during inflation $H_I=1.06\times 10^{-4}r^{1/2}M_p$   it follows that $H_I<6\times 10^{13} GeV$.

So far for the classical description of both metric and Palatini inflation. There is a striking parametric difference between
the two cases, Eq. (\ref{metrichphi}) and (\ref{palatiniphi}). The plateau begins parametrically earlier in the Palatini case, driven by the factor $\sqrt{\xi}$. And as a consequence inflation 
takes place well below the Planck scale. Asymptotically the height of the plateau is obviously the same.

\subsection{The issue of quantum corrections}\la{sec:quantum} 

The coupling $\xi$ of the non-minimal term is quite large.  This raises two concerns. The first is 
($h<<M_p$) how big the  unitarity violations in scattering amplitudes are for small Higgs field values. 
The second and most important is whether   
quantum corrections affect slow roll.  

 Elastic Higgs-Higgs scattering through one graviton exchange is the least complicated case to understand what is going on for small Higgs field. We start with the metric approach.  From reference~\cite{willenbrock} one finds for the 
 J=0 elastic amplitude with one graviton exchange in the Jordan frame that the maximum energy in  the center of mass frame is 
\be
E^2_{CM}={\pi\over 2}{M^2_p\over{(1+\xi/12)^2}}, h<<M_p.
\ee

 If  $\xi=-1/12$ there is no bound. The reason is that the amplitude vanishes in the Weyl invariant theory~\cite{callan}. For large $\xi$  the typical scale  $\L_m\equiv{M_p\over{\xi}}$  emerges for $E_{CM}$.   
 
 This result indicates that for energies well below $M_p$, at about $10^{14} GeV$, perturbative unitarity breaks down.  
 Does this mean the minimalistic view of the Standard Model 
 is not tenable in the presence of this large coupling? Not necessarily, because a strong coupling regime  for the SM could set in.
 
 The scale $\L_m$  is again retrieved  in the transition from the Jordan frame to the Einstein frame.   Eq. (\ref{dim6}) tells us that the dimension 
 six operator characteristic for the metric approach is
 \be
 {1\over{{\L^2_m}}}h^2 \partial_\m h\partial^\m h,~\mbox{with} ~\L_{m}=\sqrt{6}M_p/\xi.\non
 \ee
 This is the cut-off scale for small $h/M_p$ and is of the same order of magnitude as the unitarity limit.

A different way to come to the same conclusion is to look at how a given interaction term involving the Higgs in
the Jordan fame looks in the Einstein frame. Then there are two regimes, small Higgs values ($\xi h<<M_p$) and large Higgs values.
For the first regime the answer is what we found already for the cut-off. For the latter regime the cut-off $\L_{m}=\sqrt{6}M_p$.
This  is parametrically higher than the value of the inflationary Hubble scale (\ref{H_I}).
This was the point of the authors of reference~\cite{misha2011}: the cut-off is in principle background dependent.
But some authors have  expressed dissenting opinions ~\cite{burgess}.

\subsubsection{The Palatini case}

Let us now contrast the Palatini case with the metric case.  
We follow the same approach mentioned at the end of last section but now with the relation between the HIggs field in the 
two frames
\be
 h={M_p\over{\sqrt{\xi}}}\sinh(\sqrt{\xi}\phi/M_p).
 \ee
Take a simple term in the SM action, the Yukawa term $h \bar\psi\psi$. 
In the Einstein frame  $h$ is written in terms of the $\phi$ field  and the fermion fields transform according to their engineering dimension $\psi_E=\exp({-{3\over 2}\w})\psi$. Taking this  into account we find for the Yukawa term in the Einstein frame
\be
\exp(\w) h\bar\psi_E\psi_E={M_p\over{\sqrt{\xi}}} \tanh(\sqrt{\xi}\phi/M_p) \bar\psi_E\psi_E.\non
\ee
 
 In the small $\phi$ regime the expansion gives
 \be
 \exp(\w) h\bar\psi_E\psi_E=\phi \bar\psi_E\psi_E+{1\over{\L^2_{Pal}}}\phi^3 \bar\psi_E\psi_E+....\non
 \ee
\noindent with $\L_{Pal}={M_p\over{\sqrt{\xi}}}$  the scale where new physics or a strong coupling regime of the SM  sets in. This is parametrically less conservative than the metric case, $\L_m\sim M_p/\xi$, so $\L_{Pal}=\sqrt{6\xi}\L_m$. 

For the large $\phi$ regime one can make the reasonable guess that $\L_{Pal}\ge M_p/\sqrt{\xi}$. This in analogy
with what happens in the metric case, see the preceding subsection.

\subsubsection{Quantum corrections to the inflationary plateau}

This question is  answered by knowing to how to compute the effective potential  for the Higgs field in a reliable way.  

In the Jordan frame one can set up the effective potential in the conventional way  by a constraint in the path integral. The constraint is 
of the invariant form
\be
\d(\overline{H^\dagger H}-h^2/2).
\ee

The bar means a convenient space-time average of $H^\dagger H$.  The action is $S_J+S_m$, $S_m$ being the SM action without the Higgs sector, $S_J$ is the Jordan action introduced in the previous subsection, containing the gravitational and Higgs sector with the non-minimal term.  

The use of an invariant variable seems  obligatory because of the way one extracts the e-foldings from the potential .  

In the Einstein frame the shift invariance is  perhaps better to exploit. However it is a shift invariance in a variable that is a non-linear function of the original Higgs field.  Clearly a reliable approximation is needed~\cite{misha2011}.  

Recapitulating, we have found that for the Palatini case the cut-off is certainly parametrically larger than the inflationary Hubble constant Eq. (\ref{H_I}):
\be
\L_{Pal}\ge \sqrt{\xi} H_I.
\ee

This is certainly an encouraging result. It means that  in the inflationary phase one has not to appeal to
new degrees of freedom beyond  the SM. 

For a recent quite positive conclusion on the Palatini case  see~\cite{mcdonald}.
Gauge gravity has as mentioned more  than only scalar curvature for the gravity sector, namely the pseudo scalar curvature and the Nieh-Han invariant~\cite{misha2020}. In that context Higgs  inflation seems to be a generic phenomenon.

\section{Epilogue}

It was  in  Ann Arbor that we celebrated Tini's sixtieth birthday.  One of the contributions to the after dinner speeches was a very apt and witty rendition of his personality in the form of a poem. The title was
\be
\mbox{" The Volcano"}.\non
\ee

We have to learn to live without this force of nature.


\end{document}